\documentclass[12pt,preprint]{aastex}

\newcommand \eg     {{\it e.g., }}

\newcommand \ie     {{\it i.e.,}}
\newcommand \viz    {{\it viz.,}}

\newcommand \eq     {\,=\,}                 

\begin{document}

\title{The Structure of High Strehl Ratio Point-Spread Functions}

\author{Marshall D. Perrin\altaffilmark{1}}
	\affil{Astronomy Department\\
		University of California
		Berkeley, CA 94720}
	%
\author{Anand Sivaramakrishnan, Russell B. Makidon}
	\affil{Space Telescope Science Institute\\
		3700 San Martin Drive, Baltimore, MD 21218}
	%
\author{Ben R. Oppenheimer}
	\affil{Astrophysics Department, American Museum of Natural History\\
		Central Park West at 79th Street, New York, NY 10024}
	%
		\and
\author{James R. Graham}
	\affil{Astronomy Department\\
		University of California
		Berkeley, CA 94720}

\altaffiltext{1}{Michelson Graduate Fellow.}
\begin{abstract} 
We describe the symmetries present in the point-spread function
(PSF) of an optical system either located in space or corrected by an adaptive optics (AO) system
to Strehl ratios of about 70\% and higher.  
We present a formalism for expanding the PSF to arbitrary order
in terms of powers of the Fourier transform of the residual phase error,
over an arbitrarily shaped and apodized entrance aperture.
For traditional unapodized apertures at high Strehl ratios,
bright speckles pinned to the bright Airy rings are part of an antisymmetric
perturbation of the perfect PSF,
arising from the term that is first order in the residual phase error.
There are two symmetric second degree terms.  
One is negative at the center, and, like the first order term, is
modulated by the perfect image's field strength --- it reduces to
the Mar\'echal approximation at the center of the PSF.
The other is non-negative everywhere, zero at the image center,
and can be responsible for an extended halo ---
which limits the dynamic range of faint companion detection in
the darkest portions of the image.  In regimes where one or the other
term dominates the speckles in an image, the symmetry of the dominant
term can be exploited to reduce the effect of those speckles,
potentially by an order of magnitude or more.
We demonstrate the effects of both secondary obscuration and pupil 
apodization on the structure of residual speckles, and discuss
how these symmetries can be exploited by appropriate telescope and
instrument design, observing strategies, and filter bandwidths
to improve the dynamic range of high dynamic range AO and space-based
observations.  Finally, we show that our analysis is relevant to
high dynamic range coronagraphy.

\end{abstract}



\section{Introduction}

The direct detection of extrasolar planets, or exoplanets, has become
a major astronomical and biological focus.
The technical problems that must be overcome 
in order to image such objects near their
parent stars are formidable.  A Jovian planet is about
a million times fainter than the star at a wavelength of $1.6$\micron,
and a terrestrial planet $10^{9}$ times fainter at $0.8$\micron.
Much effort has been expended on inventing or rediscovering
shaped and apodized pupil telescopes, and developing novel
coronagraphic techniques to enable these detections.
Many of these pioneering studies assume perfect optics,
with no scattered
light, no ghost images, and a perfectly corrected wavefront.
Here we admit the possibility of imperfectly corrected wavefronts,
and investigate the structure of the point-spread function (PSF)
of a good but not perfect imaging system through an arbitrarily
shaped and apodized pupil.

We extend earlier work by \citet{Bloemhof01}, and \citet{Sivaramakrishnan02},
to show that the PSF may be expanded in an infinite Taylor-like
series, with the property that all even terms in the series are
symmetric and all odd terms are antisymmetric. Thus the sum of this
series, the overall PSF, need have no particular symmetry. However,
we show that for particular portions of parameter space (e.g.
some telescope
aperture geometry, Strehl ratio ($S$), and angular separation from the PSF
center) the sum may be dominated by one or another individual term. 
We identify two such regimes: very high Strehl ratio images ($S > 95\%$),
which are dominated almost everywhere by the antisymmetric first order
term (as described by \citet{Bloemhof03}), and moderate Strehl ratio
images ($70 \% < S < 90\%$), which are
dominated by a symmetric second-order term in the image halo,
particularly for apodized apertures. In these regimes, where speckle
formation is dominated by a single term in the PSF expansion, there
will be appreciable amounts of symmetry to the speckle patterns, which
may be exploited for significant gains in signal-to-noise (SNR) as part of
the data reduction process.

We also demonstrate theoretically and numerically that apodizing
the pupil reduces the effects of the first order antisymmetric component
of the aberrated PSF, and allows the symmetric
component to become the dominant aberration at lower Strehl ratios,
as well as 
for a larger angular extent than in the case of a completely clear aperture.
A knowledge of these PSF properties could ease the
extremely stringent demands placed on the optical quality
of a telescope dedicated to finding Jovian or terrestrial exoplanets.

In general, the PSF of an adaptively corrected optical system consists
of a diffraction limited core (e.g. an Airy function for a circular
aperture) superimposed upon an extended halo composed of numerous
"speckles" which result from the uncorrected components of the wave
front. If we characterize atmospheric turbulence using Fried's
parameter $r_0$, then there will be about $(D/r_0)^2$ speckles present in a
halo of size $D/r_0$, comparable to the size of the seeing disk in
uncorrected images. These speckles fluctuate rapidly on a timescale
of $r_0/v$ (where $v$ is the wind speed), giving speckle lifetimes
around a few hundredths of a second in practice. 

\citet{Racine99} first described the problem of speckle noise
being the dominant noise source in high Strehl ratio AO images,
by considering how clumping of uncorrected light affects the
SNR of aperture photometry near bright stars.  
\citet{Bloemhof01} developed a first-order theory of PSF structure
for symmetric, clear aperture telescopes,
and \citet{Sivaramakrishnan02} developed the complete second-order expansion
of the PSF of a well-corrected image taken with an arbitrarily shaped and
apodized pupil.  We extend the latter work to arbitrary order,
and explore the nature of various terms in the infinite series expansion
in order to determine how understanding the partially-corrected PSF
can help us design instruments to detect faint structure around stars, 
with photometry and spectroscopy of debris disks and exoplanets
as our goal.

\citet{Bloemhof01} showed that first order speckles do not appear in
dark areas of the underlying monochromatic perfectly-corrected image,
describing them as being `pinned' to the bright Airy rings on the
image.  They suggest a `speckle sweeping' strategy for the highest
Strehl ratio images, when the first order term is the dominant
aberration.  \citet{Sivaramakrishnan02} demonstrated that these pinned
speckles will still occur near the cores of images taken with
arbitrarily-shaped, unapodized entrance apertures, and that first
order pinned speckles are part of an antisymmetric perturbation on the
perfectly corrected PSF for any aperture geometry.  Knowledge of this
antisymmetry can be used to improve ways of obtaining and reducing
high Strehl ratio imaging data.  Observing strategies can be modified
to exploit this phenomenon by selecting a filter bandpass which
enables these pinned speckles to be seen, and therefore removed from
the data, thereby increasing the dynamic range of the observations
\citep{Bloemhof03}.  In this work we show that even simple apodization
of the entrance pupil suppresses the first order speckle term in
regimes accessible from either space or ground-based telescopes.


In this paper we also examine the two second degree terms in the
expansion of the PSF: \citet{Sivaramakrishnan02} identified one with
the Mar\'echal approximation or the Strehl intensity \citep{Born93},
and the other with an extended `second order halo'.  For a given
aperture, their sizes are entirely dependent on the spatial
frequencies present in the phase function.  We show that one of the
second degree terms, which we call the second order Strehl term,
dominates the first order term near the image core.

The second order halo term sets a Strehl ratio-dependent theoretical limit
on the dynamic range achievable with speckle sweeping observing strategies.
This term also sets a dynamic range limit for diffraction-limited
single-stage coronagraphy on future ExAO systems
(which will deliver Strehl ratios of $\sim$ 90\% or higher 
\citep{Makidon02}),
as well as dedicated, space-based high dynamic range imaging
telescopes such as Jovian or terrestrial planet-finding missions,
unless multi-step coronagraphic methods 
\citep{Labeyrie02, Soummer02phd} are implemented.

\section{The Full Expansion for the Partially-Corrected  PSF} \label{fullexpansion}

We briefly restate the notation of \citet{Sivaramakrishnan02}:
the telescope entrance aperture and all phase effects in a
monochromatic wavefront impinging upon the optical system are 
described by a real aperture illumination function $A(x,y)$
multiplied by a unit modulus function
$A_{\phi}(x,y) = e^{i\phi(x,y)}$.
Aperture plane coordinates are $(x,y)$ in units of the wavelength of
light, and image plane coordinates are $(\xi, \eta)$ in radians.
Deviations from a plane wave are described by a real wavefront
phase function $\phi$ which possesses a zero mean value over the
aperture plane:
    					\begin{equation} \label{zeromean}
         \int A \phi \, dx dy \,\,\Big{/} \int A \,dx dy \eq  0.
    					\end{equation}
While the choice of the phase zero point is arbitrary, selecting
the wavefront piston origin using equation (\ref{zeromean}) enables
a clean separation between the two second degree terms in our expansion,
as we will show in section \ref{second}.

The perfect optical system with no phase aberration has the
aperture illumination function $A$.

We assume that the electric field in the image plane is described by the Fourier
transform (FT) of the field in the aperture plane (\eg\ \citet{Goodman68}).  We
write the FT of a function $A$ as $a$, and the FT of $\phi$ as $\Phi$, by
changing case to indicate a transform.  The aperture illumination function with
phase aberrations is 
							\begin{equation} \label{ao_filledap}
         A_{AO} \eq  A A_{\phi},
	        					\end{equation}
with a corresponding `amplitude-spread function' (ASF) of  
$a_{AO} = a \star a_{\phi}$ (where $\star$ denotes the convolution
operation).  The ASF is proportional to to the electric field in the image plane, and is
a complex-valued function of angular image plane coordinates $(\xi, \eta)$.
The PSF of this optical system is
    											\begin{equation} \label{psf}
		p_{AO} \eq a_{AO} \,  a_{AO}^*
    											\end{equation}
(where the $^*$ operator denotes the complex conjugate).
At any location in  the pupil plane, $A_{AO}$  can be expanded
in an absolutely convergent series in $\phi$ for any
finite value of the phase function:
    											\begin{equation} \label{asf_expansion}
         A_{AO} \eq  A A_{\phi} \eq  A ( 1 + i\phi  -  \phi^2/2 + ...).
    											\end{equation}
												%

We have to Fourier transform this in order to get the ASF and PSF. In order to 
write this economically, we define the the $n$-fold convolution operator
$\star^n$ by \eg\  $x \star^3 y \equiv  x \star y \star y \star y$.
For completeness we note that $x \star^0 y \equiv x$.

Since the Fourier transform is a linear operator, we may take the
transform of the series expansion of $A_{AO}$ in equation (\ref{asf_expansion})
on a term-by-term basis, obtaining
    											\begin{equation} \label{0x01}
    a_{AO} = \sum_{k=0}^\infty  \frac{i^k}{k!} ( a \star^k \Phi),
    											\end{equation}
which possesses a complex conjugate of
    											\begin{equation} \label{0xab}
    a^*_{AO} = \sum_{k=0}^\infty  \frac{(-i)^k}{k!} ( a^* \star^k \Phi^*).
    											\end{equation}
The PSF $p_{AO} = a_{AO} \,  a_{AO}^*$ is given by the double infinite sum
	\begin{eqnarray*}
        p_{AO} &\eq&  \sum_{k=0}^\infty  \sum_{j=0}^\infty
	       \frac{i^k}{k!} (
	       a \star^k \Phi)   \frac{(-i)^j}{j!}  ( a^* \star^j \Phi^*) .
	\end{eqnarray*}
We can re-express this more usefully by grouping all terms of the
same total order in $\Phi$ and $\Phi^*$ --- that is, terms of constant $n=j+k$
(as in the binomial theorem):
	\begin{eqnarray*}
        p_{AO} 
	       &\eq&  \sum_{n=0}^\infty  \sum_{k=0}^{n}  
	       \frac{i^k (-i)^{n-k}}{k!(n-k)!}
	       ( a \star^k \Phi)  ( a^* \star^{n-k} \Phi^*) .
	\end{eqnarray*}
In other words, the $n$th order portion of the expansion of the
PSF will be given by 
    											\begin{equation} \label{0xac}
    p_n \eq  i^n \sum_{k=0}^{n}  
             \frac{(-1)^{n-k}}{k!(n-k)!}
                ( a \star^k \Phi)  ( a^* \star^{n-k} \Phi^*),
    											\end{equation}
where we have factored out the appropriate power of $i$.

Since $A$ and $\phi$ are real functions,  $a$ and $\Phi$ are
Hermitian, \ie \   their real parts are symmetric functions
of their arguments, and their imaginary parts are antisymmetric functions
(this, and all other Fourier results we use, can be found in
\citet{bracewell}).  Furthermore, multiplication, addition, subtraction
and convolution of Hermitian functions produce Hermitian functions,
so every component of each term in equation (\ref{0xac}) is Hermitian.

The sum in equation (\ref{0xac}) can further be broken into parts by
grouping pairs of terms where the power of $a$ in one matches the
power of $a^*$ in the other.
This general expansion predicts a pinned term at every order: the first and
last terms in the $n^{\rm th}$ order portion of the expansion (equation (\ref{0xac}))
contain either
only $a$ or $a^*$ as a factor --- the sum of these produces
a real contribution to the PSF which is modulated by the ASF $a$.

\subsection{Minimum Speckle Size } \label{specklesize}

Since the phase function $\phi$ (or any power of it)
never appears in the aperture illumination function in equation
(\ref{asf_expansion}) by itself, but is always multiplied by the aperture
function $A$, we only find terms such as
                $A\phi, A\phi^2,...$
in the expansion for $A_{AO}$.
Therefore the FTs of these terms, 
                $a \star \Phi, a \star^2 \Phi, a \star^3 \Phi, ...$,
are the only way we find $\Phi$ in the eventual PSF.  This means that 
all structure in the FT of the phase function $\phi$ is always
convolved with the ASF.
Even the sharpest peaks that might occur in this transform will
always appear in the PSF after such convolution.  Since the width of $a$
is just the angular resolution of the telescope, the sharpest speckles in
the PSF cannot be smaller than the angular resolution of the telescope.  
Since the expansion (\ref{asf_expansion}) converges absolutely, this
statement's veracity does not depend of the aberration being small compared
to unity, it is a general result valid for any PSF --- speckles cannot
be smaller than $\lambda/D$.  Although there are simpler ways of
demonstrating this fact, we use this argument as an exercise in tracing 
commonly-used descriptions of optical aberrations to features
in the PSF.

\subsection {Symmetry of Even and Odd Terms}

All terms of odd degree (\ie\ $p_n$, with $n$ being an odd number)
will be antisymmetric with respect to the
image origin. This follows by precisely the same logic originally applied
to the first order term: odd terms are given by $i$ times the
imaginary part of a Hermitian function, which is itself
antisymmetric. 
Likewise, all even terms must be symmetric, because they are given by
the real portions of Hermitian functions. 
What is potentially useful for increasing dynamic range of the direct
imaging described here is the fact that under different conditions 
(Strehl ratio, separation from the central star, pupil geometry and apodization),
different terms dominate the speckle noise in the image.  Knowledge of the expected
symmetry or antisymmetry can be used in instrument design, filter specification,
and data reduction to improve the SNR for the same depth of exposure.

\subsection {The Effect of Scintillation} \label{scint}

The above-derived expansion holds true for arbitrary entrance
apertures $A$. We may model scintillation as a
variation in amplitude at the telescope aperture: $A =
A_{tel}(1+\epsilon(x,y))$ where $A_{tel}$ is the telescope aperture
function in the absence of scintillation 
and $\epsilon$ is a zero-mean function representing the change in
amplitude across that aperture due to scintillation (termed 'flying
shadows' by \citet{Dravins97a}).  It is then seen
that the same expression for the expanded PSF continues to hold, with
$a$ replaced by $a \star (\delta + E)$ (where $\delta$ is a
Dirac delta function, the FT of the constant 1, and
$E$ is the FT of $\epsilon$). 
Therefore scintillation may redistribute power in a PSF compared to
the PSF without scintillation, but will always affect symmetric
speckles in a symmetric manner, and vice versa for antisymmetric
speckles.  Random changes in aperture throughput will also modify the
perfect PSF (term 0), scattering some light into the dark rings of the Airy
pattern; this will allow pinned speckles to begin appearing in these
dark rings as well.

\subsection{AO Control Radius} \label{thetaAO}

It is instructive to consider the idea of the AO control radius,
$\theta_{AO} \equiv \lambda/2d$ (where $\lambda$ is the observing wavelength,
and $d$ the inter-actuator spacing as projected back to the primary mirror--- that is,
the physical actuator spacing times the magnification factor from DM
to primary.  Here we assume that projected WFS subaperture spacing and DM actuator spacing are
comparable).
This is an angular distance in the image plane, resulting from the
inability of the deformable mirror to compensate for wavefront
aberrations beyond its spatial Nyquist frequency.
Correcting the wavefront on spatial scales up to but not beyond the
spatial frequency of $1/2d$ pulls light from an extended halo into 
the core of the PSF.  As Strehl ratios get higher, when first the
extended Mar\'echal 
($S \sim \exp{(-\sigma^2_\phi)}$)
and then the Mar\'echal 
($S \sim 1 - \sigma^2_\phi$)
approximations become
valid, the halo outside the AO control radius $\theta_{AO}$ becomes
less and less affected by improved wavefront correction within
the AO spatial frequency control band.  
When the second order expansion of the PSF is a good representation of the PSF,
correcting $\phi$ at a spatial frequency of $1/2d$ affects the PSF up to
an angle $2\theta_{AO} = \lambda/d$ from the image center, because terms which are
quadratic in the Fourier transform of the phase ($\Phi$) contain the
`doubled' spatial frequency due to the multiplication of two functions containing
signal at a spatial frequency of $\lambda/2d$.
By the time the second order terms are negligible, and the first order
expansion of the PSF is applicable, correcting the wavefront at scales
of $1/2d$ results in improvement of the PSF only within a radius of 
$\theta_{AO}$.

As a result, for Strehl ratios of about 80\% or lower, the idea of an AO control
radius holds in only an approximate sense, but as Strehl ratios climb into
the 90\% and higher range, $\theta_{AO}$ becomes a more rigorous limit on
the region of the PSF which is controlled by active or adaptive optics.
This is a direct reflection of the way $e^x$ is approximated by
a first, second, or higher order expansion in $x$.

 It has been suggested that the performance of an AO coronagraph
 can be improved by using the light which would otherwise have been
 blocked by the Lyot image plane mask for wave front sensing. Rather
 than using a dichroic to split light for WFS, a Lyot mask fabricated
 as a hole in a mirror can be used to split the light, with the
 reflected portion going to the science camera and the portion which
 passes through the hole (the core of the Airy function) going to the
 WFS. The appeal of this approach is that it could greatly reduce the
 amount of non-common-path optics between the WFS and science camera
 light paths.
 
 However, the existence of the AO control radius limits the
 effectiveness of this approach: Passing the light through a pinhole
 before sensing acts as a filter limiting the spatial frequencies
 which reach the WFS. For reasonable spot sizes, this will limit AO
 performance by preventing correction of high spatial frequency
 aberrations. For example, a Lyot spot which is $5\lambda/D$ in
 diameter will pass only spatial frequencies of 5 cycles across the
 primary diameter $D$ or lower. In this case the effective AO control radius will
 be set, not by the Nyquist frequency of the WFS, but by the maximum
 spatial frequency passed by this stop. In the case where the same
 wavelength is used for science and WFS, no useful correction is
 possible with this approach: the effective AO control radius will equal the
 spot diameter.  Speckles outside
 the Lyot spot correspond to aberrations on higher spatial
 frequencies than reach the WFS, and thus will not
 be corrected. If different wavelengths are used for WFS and science,
 then the AO control radius can be made larger than the spot diameter,
 but by a factor no greater than $\lambda_{sci}/\lambda_{WFS}$. This
 places a fundamental limit on the maximum Strehl ratio achievable
 with such systems, no matter how many WFS subapertures or DM
 actuators they may have.

However, as \citet{Macintosh02,poyneer} demonstrate, matching the
hole size and geometry to that of the wavefront sensors can improve
AO system performance. By setting the hole to the size which
corresponds to the spatial Nyquist frequency of the WFS (that is, by
making the hole radius equal to the AO control radius), this approach
will prevent aliasing of spatial frequencies in the WFS, by passing only 
those aberrations which fall within the sensing bandpass of the wavefront 
sensing system. This will result in improved WFS performance by
eliminating aliasing of uncorrectable higher spatial frequencies. However, 
the larger pinhole required for this approach is too large for
use as a Lyot spot, so this approach will still require a 
beam splitter to separate light for WFS and science.

 \section{The Second Order Expansion for the Partially-Corrected  PSF} \label{ao}

In order to demonstrate the use of our general expansion for the PSF,
we re-derive the second order expansion given by \citet{Sivaramakrishnan02}
from equation (\ref{0xac}).
This second order truncation is valid when the largest absolute value of
the phase aberration within the aperture is significantly less than unity.
Evaluating the expression in equation (\ref{0xac}) for $n=2$, we find that
	%
	\begin{eqnarray*}
    p_2 &\eq&  i^2 [ \frac{1}{2!}  ( a \star^0 \Phi)  ( a^* \star^{2} \Phi^*) \\
           &&+  \frac{-1}{1!}  ( a \star^1 \Phi)  ( a^* \star^{1} \Phi^*)  \\
           &&+ \frac{1}{2!}  ( a \star^2 \Phi)  ( a^* \star^{0} \Phi^*) ].
	\end{eqnarray*}
Combined with similar evaluations for the zeroth and first order terms, this 
yields the second order approximation to the PSF of 
												\begin{eqnarray} 
    p_{AO} &\eq&  p_0 + p_1 + p_2  \nonumber \\
           &\eq&  a  a^* \nonumber \\
        && -  i [a  (a^* \star \Phi^*)  -  a^*  (a \star \Phi)] \nonumber \\
        && +  (a \star \Phi)  (a^* \star \Phi^*)  \nonumber \\
        && -   \frac{1}{2} [a  (a^* \star \Phi^* \star \Phi^*)
                       + a^* (a \star \Phi \star \Phi)] \label{psfexp},
												\end{eqnarray}
which agrees with the result found in equation (3) of
\citet{Sivaramakrishnan02}. We now discuss the properties of each term
in this expression.

\subsection{The Perfect PSF: $p_0$} \label{zero}

In the following discussion we refer to $ p_0 \eq a \,  a^*$ as
the perfect image.
For typical clear (\ie\ unapodized) apertures, $p_0$ possesses bright
and dark rings called Airy rings.
Apodization generally fills the dark rings with light, and results in
a PSF with a greater full width at half maximum intensity, but with 
reduced diffracted light at large radii.
Using Rayleigh's theorem, we note that 
    											\begin{equation} \label{rayleigh}
		\int p_0 \, d\xi d\eta \eq  
			\int A^2 \, dx dy \eq  
				\int A_{AO}  A_{AO}^* \, dx dy.
    											\end{equation}
The physical interpretation of this is simply that whatever power enters
the optical system and is not absorbed by apodization is relayed
to the final perfect image.  Phase aberrations contribute no power
to the image, they just re-arrange the distribution of light.
By using the power theorem of Fourier theory, one can show that
the total power from the term $p_n$ contains a multiplicative
factor of 
    											\begin{equation} \label{0xff}
    i^n \sum_{k=0}^{n}  
             \frac{(-1)^{n-k}}{k!(n-k)!}.
    											\end{equation}
This factor is zero for all $n > 0$, since it is precisely
$1/n!$ times  $(1-1)^n$ (as can be seen by using the  binomial theorem
to expand the latter quantity).
The net power contained in every term except $p_0$ is therefore zero.

The perfect PSF $p_0$ is symmetric: $p_0(\xi, \eta) = p_0(-\xi, -\eta)$.
The fact that it is the power spectrum of $A$ means that it is the FT
of a symmetric, real function, specifically, the auto-correlation function
of $A$.  Therefore it possesses a symmetric real part and a zero imaginary
part, regardless of details of the aperture geometry.

\subsection{Pinned Speckles: $p_1$} \label{first}

The first order term,  
												\begin{eqnarray} 
	p_1 & \eq & - i [a  (a^* \star \Phi^*)  -  a^*  (a \star \Phi)]  \nonumber \\
	    & \eq & 2{\rm Im}[(a  (a^* \star \Phi^*))],
												\end{eqnarray}
has been discussed for arbitrary aperture geometry and apodization
by \citet{Sivaramakrishnan02}.
It was first described as being a pinned term by \citet{Bloemhof01}.
We re-iterate here that 
this term is antisymmetric:
any bright blob (due to the first order pinned term) on a bright Airy
ring must be accompanied by a corresponding dimming at the diametrically
opposite point on the ring.
Its antisymmetry forces $p_1$ to vanish at the center of the image,
\ie\ at the central peak of the PSF for typical apertures.  
This term cannot affect the Strehl ratio of the image, even though it
is the dominant cause of decreased dynamic range within a few
diffraction widths of the source in broad band AO imaging at high
Strehl ratios \citep{Sivaramakrishnan02pin}.
This pinned speckle term is entirely due to the antisymmetric
component of the function $A\phi$, because a purely symmetric
function's Fourier transform contains no imaginary component.

\subsection{The Extended Halo and the Strehl Intensity: $p_2$} \label{second}

Our choice of the phase origin which sets
the aperture-weighted mean of $\phi$ to zero
results in $a \star \Phi$ being zero at the 
image plane origin.  (We choose our image plane origin to be the
centroid of the intensity distribution, which is equivalent to 
a zero aperture-weighted mean slope of the wavefront).
This enables a natural division of the second-order contribution into two terms
with different behavior in the image plane:
												\begin{equation} 
  p_2 =   (a \star \Phi)  (a^* \star \Phi^*)  \\
         -   \frac{1}{2} [a  (a^* \star \Phi^* \star \Phi^*)
                       + a^*  (a \star \Phi \star \Phi)]  .
												\end{equation} 

The first term in the above expression is real and non-negative everywhere,
as it is of the form $zz^*$, where $z$ is a complex number.
However, because of equation (\ref{zeromean}), it is zero at the origin
(like the first order pinned speckle term),
and therefore does not affect the Strehl ratio either.
It is merely the power spectrum of the real function  $A \phi$,
a fact which ensures its symmetry about the image origin.
The slope of a Kolmogorov-spectrum atmospheric phase function comes through
to the image in this term to form the halo of
a well-corrected stellar image outside the AO control radius
$\theta_{AO} = \lambda/2d$.

This term contributes to decreased dynamic range everywhere,
because it is not modulated by the Airy pattern. 
It is the dominant term in the extended halo, as its fall-off with
radial distance from the core of the PSF is set solely
by the spatial frequencies present in the phase function.
It will place second order speckles in the dark Airy rings of a
monochromatic PSF, and will therefore set the ultimate limits
on the dynamic range of any observational speckle sweeping techniques.
We denote this term by $p_{2,\ halo}$.

The perfect PSF taken together with the {\it  second} second degree term
is the Strehl intensity \citep{Born93}.
Like the first order term $p_1$, this term is pinned to the bright
Airy rings, because it is also modulated by the size of the ASF.
Combining this with the fact that it is a second degree term
suggests that it is probably not significant outside the first
few Airy rings.  This term is also symmetric, because it is the real part
of a Hermitian function (so the entire second
degree contribution to the PSF is symmetric).
At the image center it reduces the perfect PSF by $a(0)$ 
multiplied by the
value of $(a \star \Phi \star \Phi)$ at the origin.  The former
quantity is $\int A \, dx dy$, the  latter is $\int A \phi ^2 \, dx dy$.
For a clear aperture with area $A$ this is merely $A^2 \sigma_{\phi}^2$,
which is the well-known Mar\'echal approximation
relating the Strehl ratio $S$
to the variance of the phase over the aperture, $\sigma_{\phi}^2$, 
at high Strehl ratios,
	$S = {\rm exp}(-\sigma_{\phi}^2)\  \sim\  1 - \sigma_{\phi}^2$.
For apodized apertures this yields a modified form of the Mar\'echal
approximation, since the phase variance is weighted by the apodization.
We denote this term by $p_{2,\ Strehl}$.

As mentioned in section \ref{fullexpansion}, choosing the phase
origin to ensure
that the mean of the phase (weighted by the aperture throughput)
is zero (equation (\ref{zeromean})) enables the different natures
of the two contributions to the second degree term
in the partially-corrected AO PSF to be easily distinguished.

\subsection{A Numerical Example} \label{spiefig1}

\begin{figure*}[tbp]
	\plotone{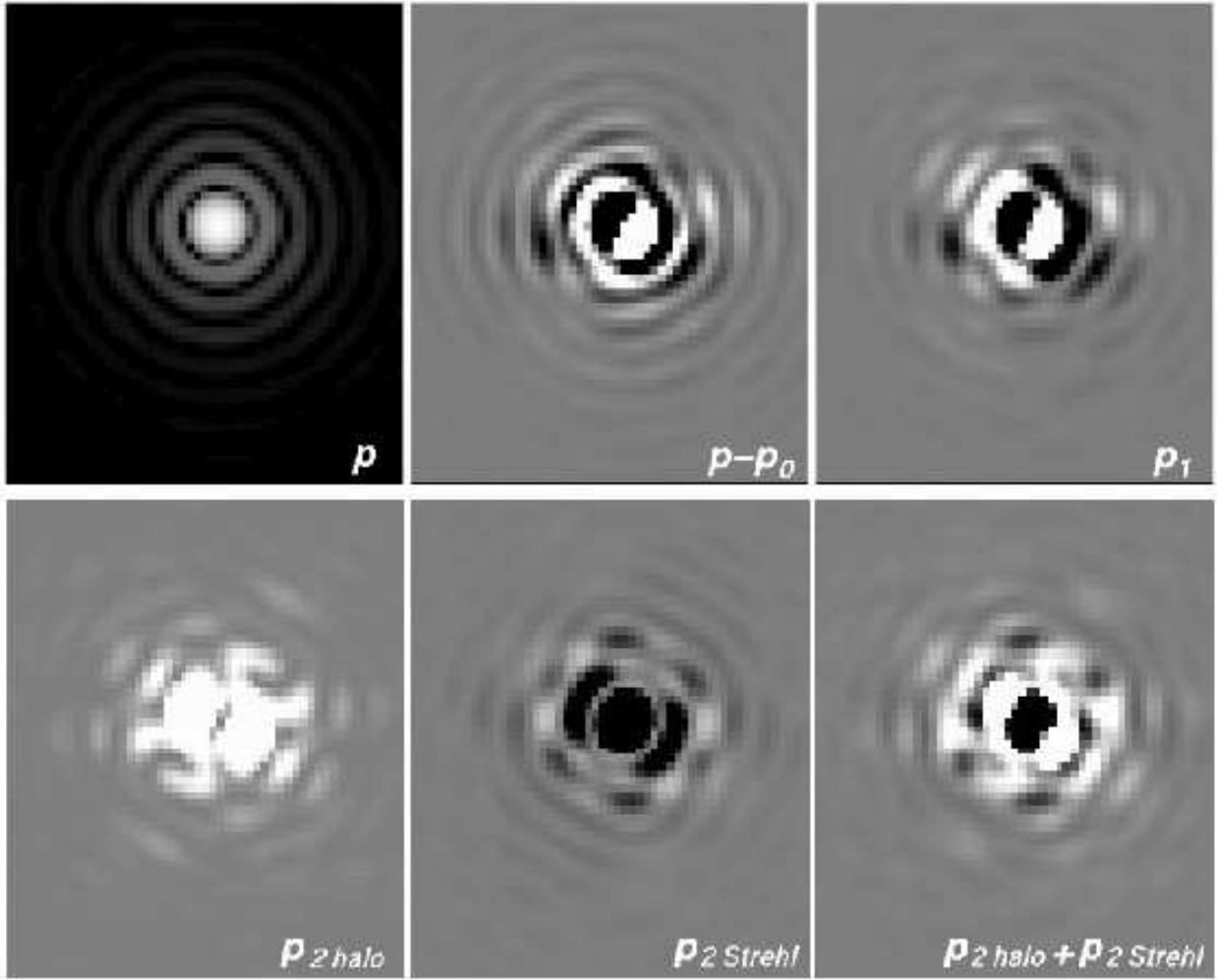}
	\caption{
	A 94\% Strehl ratio PSF, $p$, formed by a circular aperture,
	can be decomposed into individual terms in the power series in the
	Fourier transform of the phase over the aperture.
	The antisymmetry of the first order term $p_1$ and the symmetry of the 
	second order terms are apparent.  The positive halo term, $p_{2,\ halo}$,
	is zero at the image center, as is the pinned first order term $p_1$.
	At these Strehl ratios the image degradation is dominated by the pinned
	second order term $p_{2,\ Strehl}$
	(see text for details).
	}
	\label{fig:terms}
\end{figure*}

Figure \ref{fig:terms} shows a 94\% Strehl ratio PSF and its
first and second order terms.
The top left frame shows the numerical monochromatic PSF $p$
(with a logarithmic stretch) of a simulated image
formed by a circular, unobstructed aperture,
on a twice-Nyquist sampled pixel scale.
Tip-tilt errors were removed by ensuring zero mean X and Y tilts
of the incoming phase screen over the aperture, so the image center is
known {\it a priori}.
The top middle frame shows the numerical PSF with the perfect PSF
subtracted from it, ($p - p_0$), on a linear stretch.
The top right frame shows the first order pinned speckle term
$p_1$ with the same stretch as the top middle frame.
The similarity between $p_1$ and  ($p - p_0$) is obvious in the
inner two Airy rings of these two frames.

The antisymmetry of $p_1$ is also apparent in this frame. 
The total intensity in this frame is zero.
The bottom left frame shows the second order halo term $p_{2,\ halo}$ 
(on a linear stretch symmetric about zero, with black being assigned
to the minimum value of the $p_{2,\ Strehl}$ term, $-0.06$, which occurs
at the image center).
Its intensity is non-negative everywhere, and zero at the image center.
The bottom middle frame shows the pinned, second degree term, 
$p_{2,\ Strehl}$ (on the same stretch as the  $p_{2,\ halo}$ frame).
It is negative at the center, and is the only term in the 
second degree expansion of the PSF that reduces the Strehl ratio.
As noted above, this term, taken together with the perfect PSF 
(\ie\ $p_0 + p_{2,\ Strehl}$) is called the
Strehl intensity \citep{Born93}.
$p_0 + p_{2,\ Strehl}$ reproduces the Mar\'echal approximation
of classical optics theory for the image intensity at its center
at high Strehl ratios.
It contains the perfect `amplitude spread function'
(the image field strength) as a factor, so $p_{2,\ Strehl}$ could
decay more rapidly with distance from the image center than
the halo term $p_{2,\ halo}$ in ground-based AO images, depending on the
power spectrum of the phase aberration and pupil apodization.
The bottom right frame shows the sum of both second degree
terms (on the same stretch as each second degree term's images).

\placefigure{fig:terms}

\section{Numerical Exploration of Speckle Symmetry Magnitudes} \label{apodsec}
  
The relative magnitude of each term in the series expansion
of the PSF
depends on Strehl ratio, aperture shape, and aperture
apodization.
In this section we present examples of the behavior
and relative sizes of some of the lower order terms.
We simulated AO-corrected PSFs from
four circular aperture types:
circular apertures with and without central obscurations, 
and clear and apodized apertures.  For
each of four different aperture types 
(all based on the 3.6m
AEOS telescope), we ran a set of simulations, varying the number of
AO actuators from 10 to 110 across the primary, which resulted in Strehl ratios
ranging from 50\% to 98\%. 

For each simulation we generated realizations of Kolmogorov-spectrum 
turbulence phase screens, 
and, following the methods described 
in \citet{Sivaramakrishnan01}, smoothed the phase screens with a high-pass
parabolic filter to mimic the action of an AO system, \viz\  
a parabolic filter in spatial frequency space up to a Nyquist
spatial frequency corresponding to the number of actuators across
the pupil.  
This simple model of AO has been validated with fits to 70\% Strehl
ratio $K$-band data from the Palomar AO system \citep{Oppenheimer00}
and 20\% Strehl data from the 3.6m AEOS telescope on Mt. Haleakula
\citep{Makidon02}.  Scintillation effects (amplitude fluctuations across
the pupil) have not been taken into account here, but simple
calculations following the approach of \citet{Angel94} indicate these might
reduce our calculated Strehl ratio by about 5-10\% in the
$H$-band. We reiterate the point of section \ref{scint} that
scintillation will effect symmetric speckles in a symmetric manner,
and so we do not expect it to substantially change our conclusions
here.  See also \citet{ Dravins97b, Dravins98} for additional
discussion of the effects of scintillation.

We calculated the PSF of this partially-corrected wave in a perfect
imaging system by the methods of Fourier optics, using image
plane pixels of angular size  $\lambda/8D$ in order to resolve speckles clearly.
We used an input pupil 128 pixels across, in a zero-filled array 1024 pixels
on a side, thus ensuring fine sampling in our monochromatic image.
We also calculated each term in the full expansion out to fourth order
explicitly from the input phase screens and their transforms. 
The Strehl ratio of the image in Figure 1 is 94\%.

Before creating the image from the input phase screen, we subtracted
off the mean tilt of the phase function over the clear aperture.  Even
though the first order term would still retain all the properties
predicted by our theory, we felt that a large tip-tilt error obscured
the more interesting higher spatial frequency wavefront aberrations
that are found a few diffraction widths from the core of the image.

Here we describe some general properties of the results and show
representative plots of 85\% Strehl ratio images. 

\subsection{Unapodized Apertures} \label{clear}
At the currently-achievable Strehl ratio of 60\%, the second order Strehl term is 
the dominant perturbation to the perfect PSF $p_0$
on the central Airy peak, while the first order term is most important
on the first one or two Airy rings. Outside this, the second order halo term
is largest is magnitude. 
Palomar and Lick observations with Strehl ratios around 60\%
often show a three-lobed pattern of pinned speckles on the first Airy
ring \citep{Bloemhof01, Lloyd02phd}, which may be due to the dominance
of the first-order term there.

By a Strehl of 80\%, the first-order term dominates on the inner
three or four Airy rings, while the second-order halo term is responsible
for most of the residual speckle power outside of the eighth Airy ring;
the first and third order terms are down by a factor of a few relative
to the symmetric second order halo term.
In this regime data reduction techniques which take advantage of the
symmetry of the halo speckles may prove useful.

At Strehls above 90\%, the second order term decreases in magnitude 
significantly as more power moves into the diffraction limited core, 
and does not dominate until fifteen or more Airy rings out.
In this regime, data reduction
techniques exploiting the antisymmetry of the first order term should
be useful for the inner core of the image.  This effect is more pronounced
in an apodized pupil, as we discuss in section  \ref{apod}.

As Strehl ratios rise above about 98\%, the first order term increasingly
dominates throughout the entire PSF outside the core.  

\begin{figure*}[tbp]
	\plotone{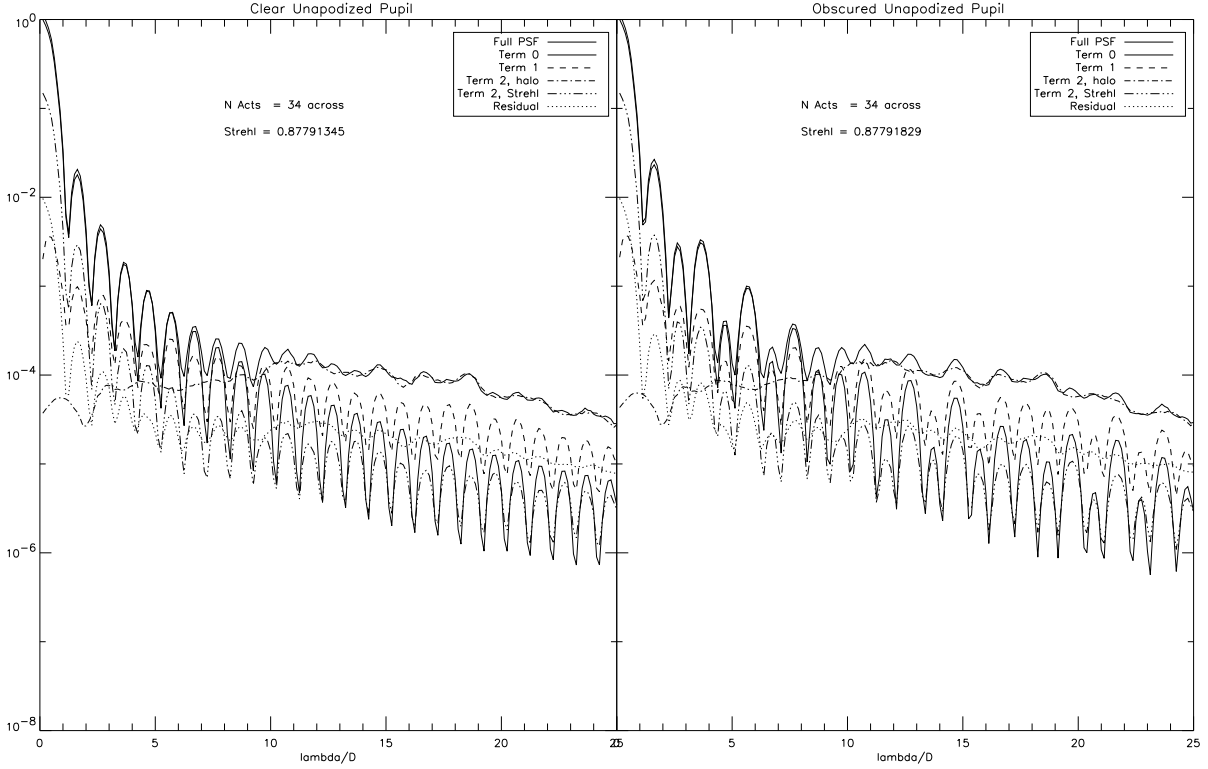}
	\caption{
    Radial profiles of the absolute value of the first few terms in the expansion
    of the PSF in terms of the Fourier transform of the phase over the aperture, 
    as well as
    the total resulting PSF, and the residual (that is, the sum of terms three
    through infinity), in simulations of the 941 actuator AO system of 
    the AEOS telescope on Haleakala.
    We plot the absolute value rather than average because the annular 
    mean of the first order antisymmetric term is identically zero everywhere.
    The left hand panel shows results for an unobscured aperture.
    Outside of about 10 $\lambda/D$ the second order halo term is clearly 
    the dominant term,
    by about an order of magnitude above both the first order term and the
    residual term (which itself is primarily due to the third and fourth order
    halo-like terms at this Strehl ratio). A secondary obscuration
    puts more power into the odd Airy rings, increasing the pinned first order
    term for those rings (right hand panel).
    This secondary obstruction decreases the separation
    between the first and second order terms, and will reduce the effectiveness
    of speckle symmetry techniques for noise reduction.
   }
	\label{fig:unapodized}
\end{figure*}

\begin{figure*}[tbp]
	\plotone{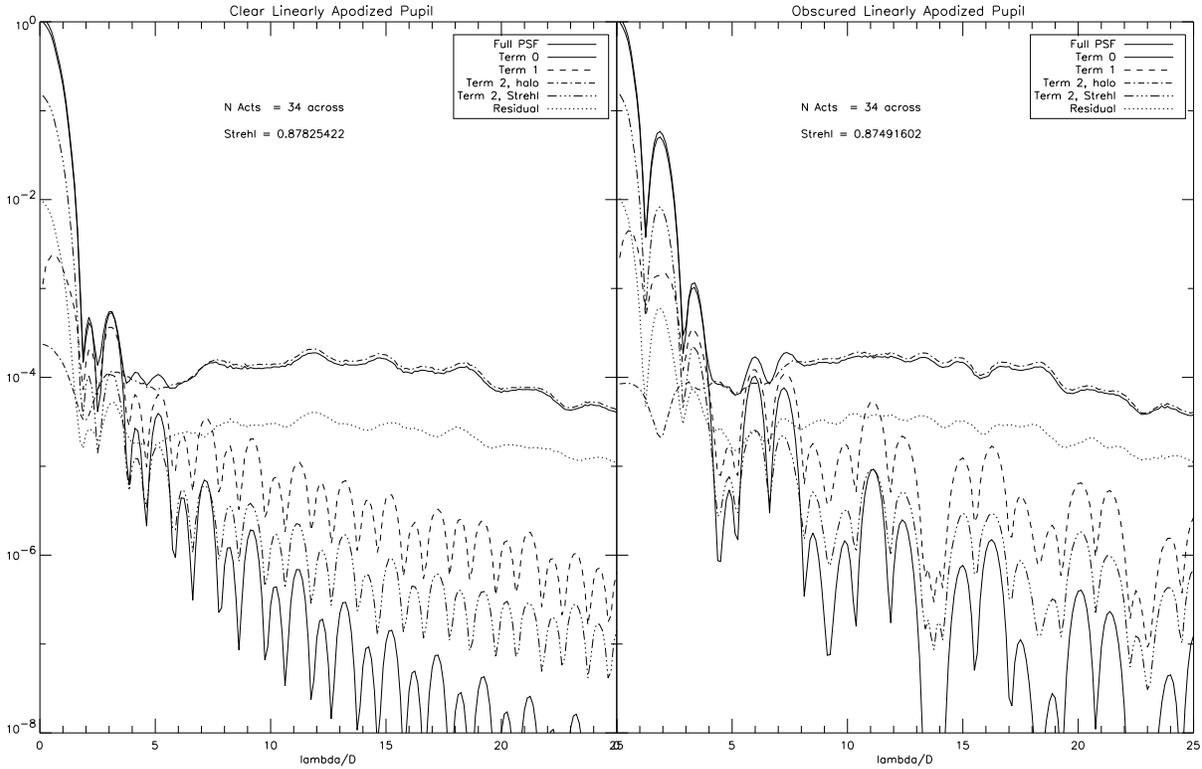}
	\caption{
    Identical to Figure \ref{fig:unapodized}, except a ``simplest-possible''
    apodization has been applied to the primary: throughput changes linearly
    from 0 at the edges of the aperture to 1 at the center. No attempt has been
    made to optimize this apodization, and it is expected that more sophisticated
    apodizations will result in better performance. However, even this simplistic
    technique dramatically reduces the perfect PSF and first order term, while
    the second-order halo is affected very little. If speckle symmetries are
    exploited to subtract off this term, the overall image halo will be reduced
    by nearly an order of magnitude. 
    }
	\label{fig:apodized}
\end{figure*}

The presence of a secondary obscuration transfers some power from the
core of the unobscured aperture PSF out to the secondary's Airy rings.
It therefore similarly
transfers power into the pinned first order term, resulting in a larger
magnitude for this term in the outer portions of the PSF, making it
closer in importance to $p_{2, Strehl}$ in the halo, as shown in
figure \ref{fig:unapodized}. This reduces the
dominance of the symmetry in that region and will therefore lower the
effectiveness of speckle symmetry subtraction for SNR gain compared to 
unobscured apertures.
                    \placefigure{fig:unapodized}
                    \placefigure{fig:apodized}

\subsection {Apodized Apertures} \label{apod}

%

The idea of telescope apodization has been explored
in several studies (\eg\ \citet{Black1980, Watson91, Nisenson01, Aime02, Soummer03}).
These pioneering efforts focussed on how apodization affects
PSFs of perfect or near-perfect optical systems.
\citet{Nisenson01} studied the effectiveness of apodized square
apertures for detection of Earth-like planets from space, arguing
that apodized square apertures had superior performance to circular
apertures along certain position angles.
\citet{Aime02} and \citet{Soummer03} derived optimized apodizations for
circular aperture telescopes equipped with Lyot coronagraphs, which 
they conclude are preferable to square apertures for high 
dynamic range imaging.
Here we inspect the role apodization plays in speckle suppression
at high Strehl ratios.

Apodization reduces the wings of the perfect unobscured aperture
PSF, so all pinned speckle terms also decrease in size.
This causes the second order halo term to become dominant at a lower 
Strehl ratio than on a comparable unapodized telescope.
Apodization is therefore beneficial not just on telescopes with
perfect optics, but on real optical systems as well.
This has been known informally for several years \citep{Ftaclas97}.
\citet{Oppenheimer03b} point out that polarization effects from thin
reflective films produce a slight natural apodization (of order a few
percent) in all concave
telescope mirrors, even those which are not intentionally apodized.
We look at residual speckle structure in high Strehl images
on pupils with a simple linear apodization (transmission changes
linearly from 0 at pupil edges to 1 at its center), in order to demonstrate 
speckle suppression caused by simple linear apodization.
Antiymmetric pinned speckles are the dominant noise source
on a 98\% Strehl image on clear apertures (see section \ref{clear}).
On a circular apodized pupil we find that between Strehl ratios of
80\% and 97\% the symmetric second order $p_{2,\ halo}$ term
is the dominant term by a factor of around 10 outside the first
few Airy rings.  This suggests that Lyot coronagraphs on such
pupils would exhibit significant speckle symmetry.
When Strehl ratios rise past  about 97\% the pinned antisymmetric
first order term, $p_1$, starts to dominate larger and larger areas around
the image core.  Thus for terrestrial planet finding from space
the linear pinned speckle term is probably the dominant cause of
speckle noise.  As \citet{Bloemhof03} suggests, observing strategies
dedicated to utilizing this property can be used under such conditions.
However, for ExAO applications from ground-based telescopes, 
apodized pupils can be beneficial even in the 75--95\% Strehl
ratio regime.

\subsection{Implications for Coronagraphy} \label{coronagraph}

\begin{figure*}[tbp]
	\epsscale{0.6}
	\plotone{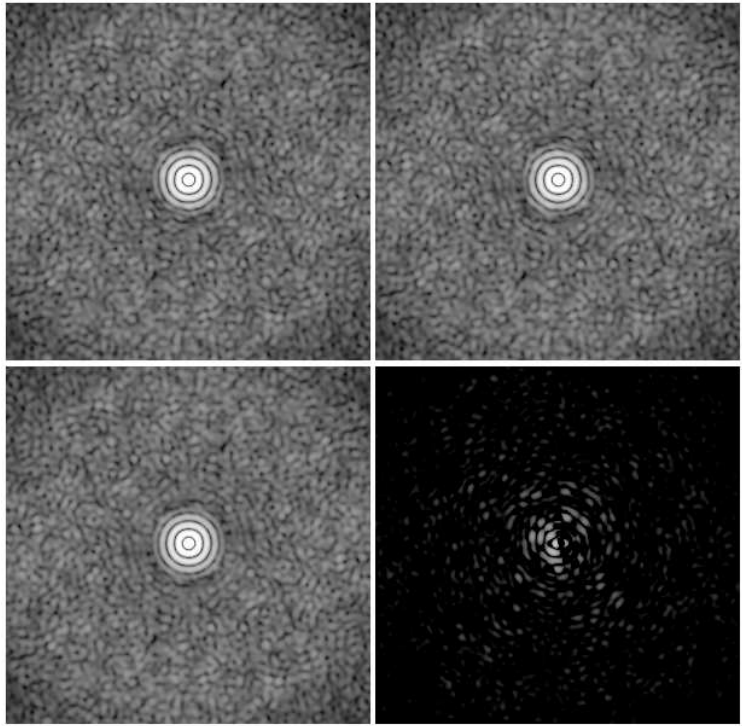}
	\caption{
    Top Left: a simulated  short exposure monochromatic coronagraphic
            image from an 8 meter telescope at the central wavelength of the $H$-band.
            Co-added PSF's from five independent realizations of Kolmogorov
			spectrum phase screens, with $r_o = 80$ cm are shown.
			The simulated AO system extrapolates Palomar AO performance to
			a 4000 actuators (71 across the primary) on Subaru,
			following \citet{Sivaramakrishnan01}.
			Scintillation has not been modelled explicitly.
            The Strehl ratio of the direct image is 98\%,
            97.6\% of the incoming intensity has been blocked by an
            optimized Lyot coronagraph with an image plane stop
            diameter of $4\lambda/D$ radians.
            All four panels are on the same logarithmic scale.  
    Top Right: the top left image rotated by $180\deg$ about its center.
    Bottom Left: the symmetric component of the original PSF,  \viz\  
            half the sum of the original image and the rotated image.
    Bottom Right: the absolute value of the antisymmetric component
            of the PSF, which is half the difference between  
            the original image and the rotated image.  
            A close inspection of this frame shows that much of the
			antisymmetric component of the speckle is generated by
			the presence of the secondary obscuration (which we
			modelled as being approximately 15\% of the primary diameter).
	}
	\label{fig:subaru}
\end{figure*}

We simulated a coronagraphic image using five independent realizations
of a Kolmogorov-spectrum phase screen incident of the aperture of an
8 meter class telescope, and a Fried 
length $r_o = 80{\rm cm}$ at the center of the $H$-band.
We model the effect of a 4000-actuator AO system with 64 actuators
across the aperture on these phase screens,
and use a hard-edged coronagraphic occulting spot $4\lambda/D$ in diameter,
with a matched, optimized Lyot pupil stop chosen in the
manner described in \citet{Sivaramakrishnan01}.  We did not include
scintillation effects, image placement error on the stop,
or scatter from mirror imperfections or dust in our simulation.
The simulated AO system produces a Strehl ratio of 98\%
(\ie\ the phase aberration alone is responsible for a 2\% Strehl hit). 
Other sources of scattered light will further degrade the Strehl
ratio.
                    \placefigure{fig:subaru}
Figure \ref{fig:subaru} shows that at higher Strehl ratios,
an understanding of image symmetry properties can improve
coronagraphic design, use, and data reduction methods.
The benefits of this understanding increase as the Strehl ratio
increases, especially in space-based imaging on very stable 
telescopes.

\subsection{Obscured Apertures} \label{obscured}

If even a small secondary obstruction is present,
the pinned terms ($p_1$ and  $p_{2,\ Strehl}$) will not die off
as quickly as the corresponding term for an unobstructed aperture.
This consideration could be relevant in methane on and methane
off bands around 1.6$\mu$m, for example, since their
fractional bandpasses (about $1/20$) produce visible Airy rings
out to several $\lambda/D$.  This argument adds weight to those
presented in \citet{Tokunaga02} for the off-axis design of a dedicated ExAO
telescope.  

An inspection of the antisymmetric component of the simulated 8-m ExAO
coronagraphic image in the lower right hand panel of
figure \ref{fig:subaru} reveals
the fact that the dominant cause of speckles within the first few
Airy rings in this image is the increased Airy ring intensity 
due to the presence of a small
secondary obstruction.
Coronagraphic performance on ground-based
ExAO systems is therefore significantly reduced by even small 
secondary obstructions.

\section {Discussion}

The expressions derived for the expansion of the point spread
function are fully general for any telescope aperture and wavefront
phase. However, our statements as to which term of that expansion
dominates under given conditions are fundamentally statistical in nature,
contingent upon the assumption of Kolmogorov turbulence which is
spatially filtered by an adaptive optics system as
described above. The presence of non-Kolmogorov atmospheric
disturbances, or of static errors in the telescope or instrumental 
optical system, may change the relative importance of the various
terms. In particular, further work is necessary
to ascertain the magnitude of the symmetric halo term for space-based
telescopes which may possess an entirely different power spectrum of
wavefront errors. 

Nonetheless, our theoretical and numerical results have a direct bearing
on current ExAO coronagraphs under construction:
the Lyot Project coronagraph on the 941-channel AO system
on the 3.6m AEOS telescope is due for first light in late
2003 \citep{Oppenheimer03}, 
and XAOPI, a coronagraph on an 8-10m class telescope \citep{Macintosh02}
is projected to start operating in 2007.
Strehl ratios of 85--95\% at a wavelength of 1.65\micron\  
are predicted for these imagers \citep{Macintosh02, Makidon02}.
AEOS and Keck non-coronagraphic infrared data taken recently will be
analyzed for the presence of features whose existence
we have predicted.

By extending the second order expansion of \citet{Sivaramakrishnan02}
to higher orders, and computing individual terms for typical realizations
of  AO-corrected atmospheric phase aberrations, we have shown when the second order
theory is applicable.  We also explain how apodization of the entrance aperture 
makes a difference to achievable dynamic range when the
images are not perfect.  Recent work \citep{Soummer03} has shown that  
optimally apodized aperture coronagraphs are more efficient than
classical Lyot coronagraphs at suppressing
the light from perfectly-corrected stellar wavefront: our analysis
demonstrates ways of determining when such apodization
will help increase dynamic range with instruments that are currently in use,
or are under construction at the moment.
Our work can also be applied to coronagraphic imaging 
to gain theoretical insight into the sensitivity of different coronagraphic
designs to residual wavefront aberrations.  Such understanding can 
affect the design of space-based telescopes dedicated to discovering 
Jovian and terrestrial exoplanets.

Our analysis of the properties of
the partially-corrected image might change the way the search space of
AO observations is modelled, thereby changing survey strategies 
and telescope and AO system design.
Clearly the fractional filter bandwidth
will limit the number of dark rings visible in an image.
The presence of a secondary obstruction will also affect the dynamic range
available to AO systems because of the more complex distribution of dark 
rings, and the slower drop-off of the perfect PSF with radius, so
first order speckles will affect a larger area than in a system
with an unobstructed pupil.  However, at Strehl ratios of
85\% and above, pupil apodization can improve dynamic range
significantly close to the on-axis object.

When reducing data, the antisymmetry of the first order speckles
is prior information that should be fed into deconvolution approaches.
Further out from the center of the image, if the symmetric halo 
dominates noise, a different data reduction strategy is recommended.
Aperture apodization and the shape of the power spectrum of the aberration
affects where in the image plane such a transition occurs.
The perfect PSF can also be folded into the deconvolution, because the
integrated power from any individual term in our expansion is zero: each term
merely redistributes power within the image.  Depending on observing
conditions and instrument performance, a knowledge of the properties
of the partially corrected PSF can be used to improve the detectability
of faint structure near bright targets.

In the case where a particular symmetry dominates the speckle pattern,
the signal to noise may be substantially improved by combining the
image with a 180 degree rotated version of itself. This is essentially
a variation on the PSF subtraction technique, in which the star serves
as its own PSF reference when suitably rotated. This avoids the
complications inherent in using a different star as a PSF reference,
at the cost of reducing the overall spatial information available.
This cost is not expected to be problematic for detection of point
sources, though it may limit the applicability of this technique for
the study of extended sources such as circumstellar disks.

In ground-based ExAO applications, our analysis is relevant to
the choice of filter bandwidth.
Our results can also be used to
optimize the length of rapid exposures
in methods that ultimately trace their origins
to the dark speckle method \citep{Labeyrie95}. 
Larger filter bandwidth smears speckles out radially, and a longer exposure
places more independent speckles all over the image.  Both these
effects cover the image with more uncorrected light.  Optimized
dark speckle techniques must balance detector read noise
considerations against photon statistics, speckle dwell times and
the effect of a finite filter bandpass (\eg\ \citet{Boccaletti01}).

The PSF properties we describe here will be even more useful for long
exposures on stable, space-based telescopes, since in that case speckles are
extremely long-lived.  A knowledge of expected speckle symmetries can be combined with
the spectral approach of \citet{Sparks02} to improve the dynamic range
limits derived in that work.
An understanding of the theory presented here can yield improved
instrument and detector design, observing strategies, and data reduction
methods in both space- and ground-based high dynamic range astronomy,
which could lead to reducing the stringency of the specifications of
the optical quality of a terrestrial or Jovian planet-finding telescope.

\acknowledgements

We are grateful to E. R. Scheinerman for the elegant proof
that the expression (\ref{0xff}) is zero.
We also wish to thank the Space Telescope Science Institute's
Research Programs Office, Visitor Program, and 
Director's Discretionary Research Fund for support, as well as the
American Museum of Natural History for its hospitality.
This work has been supported by the National Science Foundation
Science and Technology Center for Adaptive Optics, managed by the
University of California at Santa Cruz under cooperative
agreement No.  AST-9876783, and is also based upon work
supported by the National Science Foundation under
Grant No. AST-0215793. M. Perrin is supported by a NASA Michelson
Graduate Fellowship, under contract to the Jet Propulsion Laboratory
(JPL) funded by NASA. JPL is managed for NASA by the California
Institute of Technology. 

\bibliographystyle{apj}
\bibliography{ms}

\begin{thebibliography}{32}
\expandafter\ifx\csname natexlab\endcsname\relax\def\natexlab#1{#1}\fi

\bibitem[{{Aime} {et~al.}(2002){Aime}, {Soummer}, \& {Ferrari}}]{Aime02}
{Aime}, C., {Soummer}, R., \& {Ferrari}, A. 2002, \aap, 389, 334

\bibitem[{{Angel}(1994)}]{Angel94}
{Angel}, J.~R.~P. 1994, \nat, 368, 203

\bibitem[{{Black}(1980)}]{Black1980}
{Black}, D.~C. 1980, {Project Orion: A design study of a system for detecting
  extrasolar planets} (Unknown)

\bibitem[{{Bloemhof}(2003)}]{Bloemhof03}
{Bloemhof}, E.~E. 2003, \apjl, 582, L59

\bibitem[{{Bloemhof} {et~al.}(2001){Bloemhof}, {Dekany}, {Troy}, \&
  {Oppenheimer}}]{Bloemhof01}
{Bloemhof}, E.~E., {Dekany}, R.~G., {Troy}, M., \& {Oppenheimer}, B.~R. 2001,
  \apjl, 558, L71

\bibitem[{{Boccaletti} {et~al.}(2000){Boccaletti}, {Moutou}, \&
  {Abe}}]{Boccaletti01}
{Boccaletti}, A., {Moutou}, C., \& {Abe}, L. 2000, \aaps, 141, 157

\bibitem[{Born \& Wolf(1993)}]{Born93}
Born, M. \& Wolf, E. 1993, Principles of Optics, 6th ed. (Cambridge: Cambridge
  University Press)

\bibitem[{Bracewell(1986)}]{bracewell}
Bracewell, R.~N. 1986, The Fourier Transform and its Applications (London:
  McGraw Hill)

\bibitem[{{Dravins} {et~al.}(1997{\natexlab{a}}){Dravins}, {Lindegren},
  {Mezey}, \& Young}]{Dravins97a}
{Dravins}, D., {Lindegren}, L., {Mezey}, E., \& Young, A.~T.
  1997{\natexlab{a}}, \pasp, 109, 173

\bibitem[{{Dravins} {et~al.}(1997{\natexlab{b}}){Dravins}, {Lindegren},
  {Mezey}, \& Young}]{Dravins97b}
---. 1997{\natexlab{b}}, \pasp, 109, 725

\bibitem[{{Dravins} {et~al.}(1998){Dravins}, {Lindegren}, {Mezey}, \&
  Young}]{Dravins98}
---. 1998, \pasp, 110, 610

\bibitem[{Ftaclas(1997)}]{Ftaclas97}
Ftaclas, C. 1997, private communication

\bibitem[{Goodman(1968)}]{Goodman68}
Goodman, J.~W. 1968, Introduction to Fourier Optics (New York: McGraw Hill)

\bibitem[{{Labeyrie}(1995)}]{Labeyrie95}
{Labeyrie}, A. 1995, \aaps, 298, 544

\bibitem[{Labeyrie(2002)}]{Labeyrie02}
Labeyrie, A. 2002, in Astronomy with High Contrast Imaging, ed. C.~Aime \&
  R.~Soummer, EAS Publication Series

\bibitem[{Lloyd(2002)}]{Lloyd02phd}
Lloyd, J.~P. 2002, PhD thesis, University of California, Berkeley, CA

\bibitem[{{Macintosh} {et~al.}(2002){Macintosh}, {Graham}, {Ghez}, {Kalas},
  {Lloyd}, {Makidon}, {Olivier}, {Patience}, {Perrin}, {Poyneer}, {Severson},
  {Sheinis}, {Sivaramakrishnan}, {Troy}, {Wallace}, \&
  {Wilhelmsen}}]{Macintosh02}
{Macintosh}, B., {Graham}, J.~R., {Ghez}, A., {Kalas}, P., {Lloyd}, J.,
  {Makidon}, R., {Olivier}, S., {Patience}, J., {Perrin}, M., {Poyneer}, L.,
  {Severson}, S., {Sheinis}, A., {Sivaramakrishnan}, A., {Troy}, M., {Wallace},
  J., \& {Wilhelmsen}, J. 2002, American Astronomical Society Meeting, 201, 0

\bibitem[{Makidon {et~al.}(2002)Makidon, Sivaramakrishnan, {Roberts, Jr.},
  Oppenheimer, \& Graham}]{Makidon02}
Makidon, R.~B., Sivaramakrishnan, A., {Roberts, Jr.}, L.~C., Oppenheimer,
  B.~R., \& Graham, J.~R. 2002, in Proc. SPIE, Vol. 4860, High-Contrast Imaging
  for Exo-planet Detection, ed. A.~B. Schultz \& R.~G. Lyon

\bibitem[{{Nisenson} \& {Papaliolios}(2001)}]{Nisenson01}
{Nisenson}, P. \& {Papaliolios}, C. 2001, \apjl, 548, L201

\bibitem[{Oppenheimer \& Breckinridge(2003)}]{Oppenheimer03b}
Oppenheimer, B. \& Breckinridge, J. 2003, Submitted, Ap.J.

\bibitem[{Oppenheimer {et~al.}(2000)Oppenheimer, Dekany, Hayward, Brandl, Troy,
  \& Bloemhof}]{Oppenheimer00}
Oppenheimer, B.~R., Dekany, R.~G., Hayward, T.~L., Brandl, B., Troy, M., \&
  Bloemhof, E.~E. 2000, in Proc. SPIE, Vol. 4007, Adaptive Optical Systems
  Technologies, ed. P.~L. Wizinowich, 899--905

\bibitem[{Oppenheimer {et~al.}(2003)Oppenheimer, Sivaramakrishnan, \&
  Makidon}]{Oppenheimer03}
Oppenheimer, B.~R., Sivaramakrishnan, A., \& Makidon, R.~B. 2003, in The Future
  of Small Telescopes (Kluwer Academic, Dordrecht)

\bibitem[{Poyneer \& Macintosh(2003)}]{poyneer}
Poyneer, L. \& Macintosh, B. 2003, Lawrence Livermore National Lab document

\bibitem[{{Racine} {et~al.}(1999){Racine}, {Walker}, {Nadeau}, {Doyon}, \&
  {Marois}}]{Racine99}
{Racine}, R.~., {Walker}, G.~A.~H., {Nadeau}, D., {Doyon}, R.~., \& {Marois},
  C. 1999, \pasp, 111, 587

\bibitem[{Sivaramakrishnan {et~al.}(2002)Sivaramakrishnan, Hodge, Makidon,
  Perrin, Lloyd, \& Bloemhof}]{Sivaramakrishnan02pin}
Sivaramakrishnan, A., Hodge, P.~E., Makidon, R.~B., Perrin, M.~D., Lloyd,
  J.~P., \& Bloemhof, E.~E. 2002, in Proc. SPIE, Vol. 4860, High-Contrast
  Imaging for Exo-Planet Detection, ed. R.~G. Lyon \& A.~B. Schultz

\bibitem[{{Sivaramakrishnan} {et~al.}(2001){Sivaramakrishnan}, {Koresko},
  {Makidon}, {Berkefeld}, \& {Kuchner}}]{Sivaramakrishnan01}
{Sivaramakrishnan}, A., {Koresko}, C.~D., {Makidon}, R.~B., {Berkefeld}, T., \&
  {Kuchner}, M.~J. 2001, \apj, 552, 397

\bibitem[{{Sivaramakrishnan} {et~al.}(2002){Sivaramakrishnan}, {Lloyd},
  {Hodge}, \& {Macintosh}}]{Sivaramakrishnan02}
{Sivaramakrishnan}, A., {Lloyd}, J.~P., {Hodge}, P.~E., \& {Macintosh}, B.~A.
  2002, \apjl, 581, L59

\bibitem[{Soummer(2002)}]{Soummer02phd}
Soummer, R. 2002, PhD thesis, Universit\'e de Nice -- Sophia-Antipolis, Nice
  Sophia-Antipolis address

\bibitem[{Soummer {et~al.}(2003)Soummer, Aime, \& Falloon}]{Soummer03}
Soummer, R., Aime, C., \& Falloon, P.~E. 2003, \aaps, 397, 1161

\bibitem[{{Sparks} \& {Ford}(2002)}]{Sparks02}
{Sparks}, W. \& {Ford}, H.~C. 2002, \apj, 578, 543

\bibitem[{Tokunaga {et~al.}(2002)Tokunaga, Ftaclas, Kuhn, \&
  Baudoz}]{Tokunaga02}
Tokunaga, A.~T., Ftaclas, C., Kuhn, J.~R., \& Baudoz, P. 2002, in Brown Dwarfs,
  ed. E.~Martin, IAU Symposium {\em 211}

\bibitem[{Watson {et~al.}(1991)Watson, {Mills}, {Gaiser}, \& Diner}]{Watson91}
Watson, S.~M., {Mills}, J.~P., {Gaiser}, S.~L., \& Diner, D.~J. 1991, Applied
  Optics-OT, 30, 3253

\end{thebibliography}

\newpage

\clearpage

\clearpage

\clearpage

\end{document}